\documentclass[prb,twocolumn,showpacs,amsmath,amssymb,superscriptaddress]{revtex4-1}

\usepackage{float}
\usepackage{graphicx}
\usepackage{dcolumn}
\usepackage{mathtools}
\usepackage{multirow}
\usepackage{bm}
\usepackage{hyperref}
\hypersetup{pdfnewwindow=true, colorlinks=true, linkcolor=blue, anchorcolor=blue, citecolor=blue, filecolor=blue, menucolor=blue, urlcolor=blue}

\begin{document}

\title{Large cross-polarized Raman signal in CrI$_3$, A first-principles study}

\author{Ming Lei}
\email{mlei012@ucr.edu}
\affiliation{Chemical and Environmental Engineering, University of California Riverside, CA 92521, USA}
\author{Sinisa Coh}
\affiliation{Materials Science and Mechanical Engineering, University of California Riverside, CA 92521, USA}

\date{\today}

\begin{abstract}
We find unusually large cross-polarized (and anti-symmetric) Raman signature of A$_{\rm g}$ phonon mode in CrI$_3$, in agreement with experiments.  The signal is present only when the following three effects are considered in concert: ferromagnetism on Cr atoms, spin-orbit interaction, and resonant effects.  Somewhat surprisingly, we find that the relevant spin-orbit interaction potential originates from iodine atoms, despite magnetism being mostly on chromium atoms.  We analyze the Raman signature as a function of magnetic order, the direction of the magnetic moment, energy and polarization of light used for Raman scattering, as well as carrier lifetime.  In addition to a strong cross-polarized Raman signal, we also find unusually strong phonon modulated magneto-optical Kerr effect (MOKE) in CrI$_3$.
\end{abstract}

\maketitle

\section{Introduction}

A two-dimensional magnet CrI$_3$ has attracted attention due to its potential applications in magneto-optoelectronic,\cite{huang2017layer, seyler2018ligand, wu2019physical} spintronic devices,\cite{Zhonge1603113, jiang2018electric, jiang2019spin, jin2020evidence, PhysRevB.102.081117} and data storage.\cite{wang2018very} Polarized Raman spectroscopy is a powerful and commonly used experimental tool to probe the magnetic\cite{PhysRevX.10.011075, cenker2020direct, klein2019enhancement, doi:10.1063/1.1714194, 9b04634,mccreary2019distinct,huang2020tuning,li2019pressure,guo2020layer,C8CP03599G, PhysRevB.98.085406} and structural\cite{PhysRevB.98.104307, admi.201901410,
PATIL2020100174,doi:10.1021/acs.nanolett.8b01131} properties of CrI$_3$ as it is both non-invasive and extremely sensitive on the electronic and phononic structure of the material.

At low temperatures, below 210--220~K,\cite{cm504242t} bulk CrI$_3$ crystallizes in the rhombohedral space group R$\overline{3}1'$.  Point symmetries present in this phase are three-fold rotation axis, inversion symmetry, time-reversal symmetry ($1'$), and any combination of these symmetries taken together.  Cr atoms are at Wyckoff 6c sites while I atoms are at Wyckoff 18f sites. The structure units are edge-shared CrI$_6$ octahedra. When the temperature is reduced further, below 68~K, the bulk CrI$_3$ orders ferromagnetically \cite{doi:10.1063/1.1714194} which breaks the time-reversal symmetry ($1'$). Therefore the symmetry of the system is reduced,
$$\rm{from \ R}\overline{3}1' {\rm \ to \ R}\overline{3}.$$
In both cases, by symmetry, one expects four A$_{\rm g}$ Raman active modes in CrI$_3$,\footnote{The Wyckoff 6c orbit of Cr has one free parameter, while 18f orbit of I atom has three free parameters.  Therefore, there must be in total four symmetry-preserving A$_{\rm g}$ modes.} however the corresponding Raman tensors are different in the magnetic and non-magnetic states. In particular, the symmetry of the non-magnetic state (R$\overline{3}1'$ symmetry) imposes that the Raman tensor has the form,\cite{itd}
\begin{align}
{\rm (nonmagnetic)} \quad
{\rm R}\overline{3}1' \quad \rightarrow \quad
\begin{pmatrix*}[l]
a & 0 & 0 \\
0 & a & 0 \\
0 & 0 & c
\end{pmatrix*}_{{\rm A}_{\rm g}}
\label{eq:tensor_nonmag}
\end{align}
Therefore the A$_{\rm g}$ modes in the non-magnetic state of CrI$_3$ don't have cross-polarized $I_{xy}$ Raman intensity, but have parallel-polarized $I_{xx}$ intensity ($I_{xx}\sim |a|^2$). In other words, polarization of incoming and scattered light to CrI$_3$ is the same. However, in the ferromagnetic (FM) state of CrI$_3$ the symmetry dictates that the Raman tensor has the following form,\cite{itd}
\begin{align}
{\rm (ferromagnetic)} \quad
{\rm R}\overline{3} \quad \rightarrow \quad
\begin{pmatrix*}[l]
\phantom{-}a & b & 0 \\
-b & a & 0 \\
\phantom{-}0 & 0 & c
\end{pmatrix*}_{{\rm A}_{\rm g}}
\label{eq:tensor_mag}
\end{align}
The presence of the anti-symmetric term $b$ in the Raman tensor contributes to the Raman scattering intensity $I_{xy}$ ($\sim |b|^2$) in the cross-polarized configuration, as shown in Ref.~\onlinecite{Cracknell_1969}.

The form of Raman tensor from Eq.~\ref{eq:tensor_mag} is consistent with experiment.  Furthermore, the dominant A$_{\rm g}$ mode shows an unusually high intensity in the cross-polarized ($I_{xy}$) configuration at 128~cm$^{-1}$. However, the measured intensity ratio
\begin{align}
R = \frac{I_{xy}}{I_{xx}}
\end{align}
greatly varies from one experiment to the other. For example, in the case of CrI$_3$ monolayer in the FM state, Refs.~\onlinecite{9b04634},~\onlinecite{mccreary2019distinct},~\onlinecite{huang2020tuning}
find that the intensity ratio $R$ is 1.5, 0.5, and 0.65, respectively. While all three experiments use 633~nm laser, the temperatures are slightly different, 1.7~K, 9~K, and 15~K.

With more layers of CrI$_3$ the measured ratio $R$ tends to decrease.~\cite{9b04634}  For example, Ref.~\onlinecite{PhysRevB.98.104307}
(532~nm, 100~K) and Ref.~\onlinecite{doi:10.1021/acs.nanolett.8b01131} (room temperature) both report that in bulk CrI$_3$ the ratio $R$ is about 0.14.  Finally, both Ref.~\onlinecite{li2019pressure} (bulk, 633~nm, 300~K and 90~K) and Ref.~\onlinecite{guo2020layer} (2-4~layers and bulk, 633~nm, 10~K) find that $R \approx 0$.

One possible explanation for such a wide range of reported values of $R$ is that $R$ depends on the experimental conditions such as the energy of the incoming photons, quality of the sample, temperature relative to the FM phase transition, and so on. For example, the temperature in Refs.~\onlinecite{PhysRevB.98.104307, li2019pressure,doi:10.1021/acs.nanolett.8b01131} is above the Curie temperature (61~K), so we do not expect large $R$. Furthermore, Ref.~\onlinecite{PhysRevX.10.011075} found a mixed state of layered anti-ferromagnetic state (AFM) on the surface and FM in the bulk. The presence of AFM state should reduce $R$ to near zero, as in AFM state time-reversal followed by spatial inversion is a symmetry of the system, which is enough to enforce $R=0$.  Given the uncertainties in the experimentally determined value of $R$,  we set out to compute the Raman intensity of 128~cm$^{-1}$ A$_{\rm g}$ mode from first-principles. 

Our first-principles calculations find large cross-polarized Raman signature $I_{xy}$ of A$_{\rm g}$ phonon mode in CrI$_3$.  The intensity of $I_{xy}$ Raman signal is strongly dependent on the frequency of the incoming light.  We find that the  $I_{xy}$ polarized Raman scattering is driven by ferromagnetism, spin-orbit interaction, and resonance effects. Moreover, the relevant spin-orbit interaction potential comes from iodine atoms, rather than the magnetic Cr atoms.  This finding is consistent with earlier studies that also report importance of spin-orbit interaction on the nominally non-magnetic atom.\cite{PhysRevB.99.104432, Tartagliaeabb9379, Lado_2017, D0TC01322F, Mukherjee_2019}

While in this work we focus on only one A$_{\rm g}$ mode, the one near 128~cm$^{-1}$,  by symmetry one expects three other A$_{\rm g}$ modes in CrI$_3$. However, two of the three modes (expected to be located at around 100 and 200~cm$^{-1}$ respectively) are likely too weak to be detected experimentally.\cite{PhysRevB.98.104307,9b04634,huang2020tuning}
The remaining A$_{\rm g}$ mode has been measured at Raman shift of around 75~cm$^{-1}$.  The measured ratio between $I_{xy}$ and $I_{xx}$ polarized intensities for this mode are 0.5,\cite{9b04634} 0.3\cite{huang2020tuning} and 0.13 \cite{PhysRevB.98.104307}, respectively. 

Finally, we note that non-zero $I_{xy}$ Raman signature of A$_{\rm g}$ modes could also, in principle, originate from a hypothetical breaking of a three-fold rotation axis, and not necessarily from breaking of the time-reversal symmetry.\footnote{We note that a high-temperature phase of CrI$_3$ has $C2/m$ monoclinic space group, and thus does not contain a 3-fold axis.  Nevertheless, the $C2/m$ group contains a 2-fold rotation axis, oriented within the CrI$_3$ plane, as well as a mirror symmetry.  Either of these symmetries require $I_{xy}=0$.  Here we choose convention in which the $y$-axis is pointing along the monoclinic 2-fold rotation axis. Finally, while $I_{xy}=0$ is zero, the $I_{xz}$ component is non-zero for the A$_{\rm g}$ mode in the monoclinic phase.}  However, there are other characteristics of the Raman signal, not only the fact that $I_{xy}$ is non-zero, that can help distinguish between a broken three-fold rotation axis and a broken time-reversal symmetry.  These differences arise from the fact that the time-reversal symmetry induces anti-symmetric off-diagonal component of the Raman tensor $b$ while breaking of the three-fold rotation axis induces symmetric off-diagonal components,
\begin{align}
\begin{pmatrix*}[l]
. & b & . \\
b & . & . \\
. & . & . \\
\end{pmatrix*}_{{\rm A}_{\rm g}} .
\label{eq:tensor_no3}
\end{align}
(Only two matrix elements are shown for clarity.) The first qualitative difference between the anti-symmetric (Eq.~\ref{eq:tensor_mag}) and the symmetric (Eq.~\ref{eq:tensor_no3}) off-diagonal Raman tensor is that in the former the Raman intensity for left and right circularly polarized light are different from each other.  The second difference comes from the fact that in the case of the time-reversal symmetry breaking $b$ is induced by the magnetic moment $M$, so changing the sign of $M$ will also change the sign of $b$.  Therefore, for example, the ratio of left and right circularly polarized Raman intensity will be inversed if the direction of the magnetic moment is changed.  Third, there are also other qualitative differences, even in the case of linearly polarized Raman signal where polarization angle between incoming and outgoing light is not exactly 90$^{\circ}$.  All three of these qualitative signatures of the time-reversal symmetry breaking were found experimentally.\cite{huang2020tuning}

This paper is organized as follows: in Sec.~\ref{sec:approach} we describe our approach, in Sec.~\ref{sec:results} we present our results, analysis is done in Sec.~\ref{sec:analysis}, and we conclude in Sec.~\ref{sec:conclusion}.

\section{Approach}
\label{sec:approach}

Now we discuss our approach to compute the Raman intensity in CrI$_3$. By symmetry the cross-polarized term $b$ appearing in Eq.~\ref{eq:tensor_mag} is non-zero whenever ferromagnetic order is present. However, the symmetry alone can not tell us the magnitude of $b$.  Formally, the Raman scattering intensity for light with incoming frequency $\omega_{\rm I}$ is given by the derivative of the susceptibility tensor $\chi_{\alpha \beta} (\omega_{\rm I})$ with respect to atomic displacements $u$.\cite{peter2010fundamentals} In particular, the intensity $I_{xx}$ is given as,
\begin{align}
I_{xx} (\omega_{\rm I}) \sim  \left\vert a(\omega_{\rm I}) \right\vert^2
\omega_{\rm I}^4
\sim
\left \vert \frac{\partial \chi_{xx} (\omega_{\rm I}) }{\partial u} \right\vert^2
\omega_{\rm I}^4,
\label{eq:intensity_xx}
\end{align}
while for the cross-polarized intensity $I_{xy}$ we have,
\begin{align}
I_{xy} (\omega_{\rm I}) \sim  \left\vert b(\omega_{\rm I}) \right\vert^2
\omega_{\rm I}^4
\sim
\left \vert \frac{\partial \chi_{xy} (\omega_{\rm I}) }{\partial u} \right\vert^2
\omega_{\rm I}^4. \label{eq:intensity_xy}
\end{align}
In our approach we neglect the energy of a phonon, as it is small relative to other energy scales in the problem.

Three conditions need to be met simultaneously to arrive at non-zero $b$ for CrI$_3$, and thus non-zero cross-polarized intensity $I_{xy}$.  First, as discussed earlier, CrI$_3$ must be in a state with broken time-reversal symmetry, such as ferromagnetic state.  Second, one needs to consider spin-orbit interaction, as otherwise electronic wavefunctions are insensitive on the overall direction of the magnetic moment, and $\chi_{xy}=0$ regardless of atomic displacements ($u$), so $\frac{\partial \chi_{xy}}{\partial u}$ must be $0$ as well.  Third, we need to consider Raman scattering in the non-resonant limit, where the frequency of light is non-zero ($\omega_{\rm I} > 0$).  As is known from the theory of the anomalous Hall effect, $\chi_{xy} (\omega_{\rm I}=0)$ is quantized for an insulator, so its derivative must vanish.  Therefore in the limit of zero frequency of incoming light ($\omega_{\rm I} \rightarrow 0$) coefficient $b$ must vanish in any insulator.

Earlier first-principles calculations of CrI$_3$ did not take into account two of three effects listed here (spin-orbit interaction and resonance), which is why theoretical studies Ref.~\onlinecite{C8CP03599G, PhysRevB.98.085406} report $b=0$ and therefore $R=0$. We report here the first calculations of the Raman tensor in CrI$_3$ that includes both the spin-orbit interaction and the resonance effects. We will discuss in more detail the role of each of the three conditions in Sec.~\ref{sec:analysis}.  

\subsection{Computational details}
\label{sec:comp}

For density functional theory calculations we use the Quantum Espresso package.\cite{Giannozzi_2009} We use the local density approximation (LDA) along with the optimized norm-conserving Vanderbilt (ONCV) pseudopotentials which include the spin-orbit effect.\cite{PhysRevB.88.085117} These pseudopotentials describe the valence electrons 3s3p3d4s in Cr and 5s5p in I. In order to obtain sufficient precision, we cutoff the plane wave basis for the wavefunction at 80~Ry, and sample the electron's Brillouin zone on a 5$\times$5$\times$5 Monkhorst-Pack grid. Given this coarse k-grid we use Wannier interpolation to calculate the susceptibility tensor $\chi$ on a grid with 30$\times$30$\times$30 points.\cite{MOSTOFI20142309} We use experimental crystal structure from Ref.~\onlinecite{PhysRevB.97.014420} in our calculations. For the monolayer calculation, the thickness of the unit cell perpendicular to the monolayer is 20~\AA. We computed magnetic moments by integrating magnetic moment density inside a sphere of radius 1.1~\AA\ centered on an atom.

Bulk CrI$_3$ has significant excitonic effects, as computed in Ref.~\onlinecite{wu2019physical} from the Bethe-Salpeter equation approach (GW-BSE).  The exciton absorption was calculated to start around 1.7~eV, while the bulk continuum starts above 2.0~eV.  In the present work, we focus mostly on the Raman intensity ratio $R$, so we computed the susceptibility tensor $\chi_{\alpha \beta} (\omega_{\rm I})$ and its derivative from the approach that is less computationally demanding than the GW-BSE method.  We used the Kubo formula\cite{JPSJ.12.570} in the random phase approximation (RPA) using the density functional theory (DFT) computed eigenstates and eigenenergies,
\begin{equation}
\begin{aligned}
&\sigma_{\alpha \beta} (\omega_{\rm I})=
\\
&
\frac{ie^2\hbar}{V N_k}
\sum_{\textbf{k}}\sum_{n m}\frac{f_{m \textbf{k}} - f_{n\textbf{k}}}{\varepsilon_{m\textbf{k}} - \varepsilon_{n\textbf{k} }}\frac{\mathinner{\langle \phi_{n\textbf{k}}|} v_{\alpha} \mathinner{|\phi_{m\textbf{k}}\rangle\mathinner{\langle \phi_{m\textbf{k}}|} v_{\beta} \mathinner{|\phi_{n\textbf{k}}\rangle}}}{ \varepsilon_{m\textbf{k}} - \varepsilon_{n\textbf{k}} - \hbar\omega_{\rm I} - i \delta/2}. \label{Kubo}
\end{aligned}
\end{equation}
$\alpha$ and $\beta$ denote Cartesian directions. $V$ is the cell volume, $N_k$ is the number of k-points, and $f_{n\textbf{k}}$ is the Fermi-Dirac distribution function. $\omega_{\rm I}$ is the optical frequency.  We assume a constant lifetime of electronic states $1/\delta$ in the Kubo formula.  Later we discuss our choice for parameter $\delta$. Since DFT based RPA approach is not meant to give the correct band gap in the calculation, we applied a rigid shift to the computed spectrum to reproduce the gap onset as computed within GW-BSE.  Therefore, we expect that onset of the Raman signature will be well reproduced in our calculation, but that there will be rearrangement of oscillator weight, especially in the reduced screening environment of a CrI$_3$ monolayer, that will not be present in our RPA approach.  Therefore, while two-particle effects will redistribute some of the Raman signatures above the gap,  we expect that the overall trends of the computed Raman intensities will be well reproduced, especially in bulk CrI$_3$.

To obtain the Raman intensity we first calculate the susceptibility tensor components $\chi_{xx} (\omega_{\rm I})$ and $\chi_{xy} (\omega_{\rm I})$ in the ground state, without any atomic displacements. Next, we displace atoms by a small finite amount and recalculate the susceptibility tensors. The calculated difference of the susceptibility tensors then gives us $\frac{\partial \chi (\omega_{\rm I}) }{\partial u}$. We obtain the same numerical value of derivative if we displace atoms by either 0.01~$\rm \AA$ or 0.02~$\rm \AA$.  This finding confirms the linearity of $\Delta \chi$ on atom displacement.

\section{Results}
\label{sec:results}

In this study we focus on the A$_{\rm g}$ mode at frequency close to 128~cm$^{-1}$.  While there are four A$_{\rm g}$ degrees of freedom contributing to the 128~cm$^{-1}$ mode, the dominant motion is coming from the movement of iodine atoms along the 3-fold rhombohedral axis ($z$-axis).\cite{9b04634,PhysRevB.98.104307}  Nevertheless, we also computed for completeness Raman signal of remaining three types of motions within the A$_{\rm g}$ manifold.  One of these modes contributes to the displacement of Cr atoms along the $\pm z$ axis while the other two consist of in-plane ($x$--$y$) displacement of I atoms.
We measured the relative importance of all four A$_{\rm g}$ degrees of freedom by computing,
\begin{align}
\frac{1}{\omega_2-\omega_1}
\int_{\omega_1}^{\omega_2} \left\vert \frac{\partial \chi_{xy}(\omega_{\rm I})}{\partial u} \right\vert d \omega_{\rm I}
\label{eq:averageschi}
\end{align}
for each mode.  The average here is performed over the optical region (between 1.0 and 2.5~eV) to eliminate oscillations of $\frac{\partial \chi_{xy}(\omega_{\rm I})}{\partial u}$ in frequency.  The computed average is 11~${\rm \AA}^{-1}$ for the dominant A$_{\rm g}$ mode corresponding to the displacement of I atoms along $\pm z$ axis. The average for displacement of Cr atoms along the $\pm z$ axis is only 1.2~${\rm \AA}^{-1}$, while averages for displacements of I atoms in the $x$--$y$ plane are even smaller (0.4~${\rm \AA}^{-1}$ and 0.3~${\rm \AA}^{-1}$). Since the 128~cm$^{-1}$ is dominated by the movement of I atoms along the $z$-axis, and that the Raman intensity is proportional to the square of $| \frac{\partial \chi_{xy}}{\partial u}|$, we are justified in focusing only on the $\pm z$ displacement of iodine atoms in our analysis.

\subsection{Bulk FM CrI$_3$}

First we discuss the calculated Raman intensities $I_{xx}$ and $I_{xy}$ of the 128~cm$^{-1}$ mode in bulk CrI$_3$, as shown in Fig.~\ref{fig:intensity}. Raw data used to obtain these intensities are provided in the supplement.\cite{} The DFT computed bulk bandgap is 0.79~eV, and the horizontal axis is rigidly shifted by $0.71$~eV so that it matches the GW-BSE calculated optical band gap of 1.5~eV.\cite{wu2019physical} While the value of electron lifetime is not known in CrI$_3$ we used in this calculation a conservative value of $\delta = 0.1$~eV, same as that in graphite.\cite{PhysRevB.69.205416,Kumar_Gudelli_2019} We note that in the experiment this value might depend on the sample preparation.  Later in Sec.~\ref{sec:lifetime} we show that for an even larger $\delta$ the calculated ratio $R$ has nearly the same shape, but is overall smoother as a function of $\omega_{\rm I}$.

\begin{figure}[h]
\centerline{\includegraphics{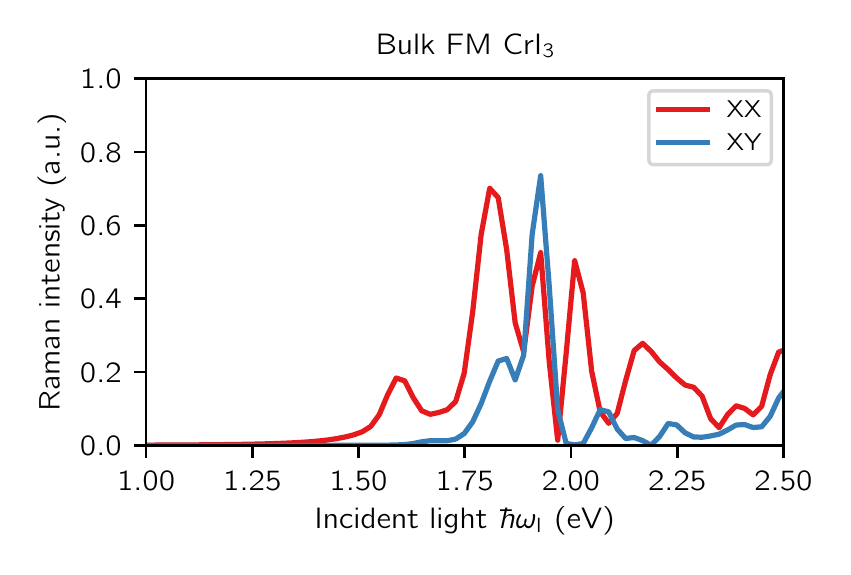}}
\caption{\label{fig:intensity} The calculated Raman intensity of $A_{\rm g}$ mode in bulk ferromagnetic CrI$_3$. The photon energy on the horizontal axis is shifted to reproduce the GW-BSE calculated optical bandgap onset.}
\end{figure}

\begin{figure}[h]
\centerline{\includegraphics{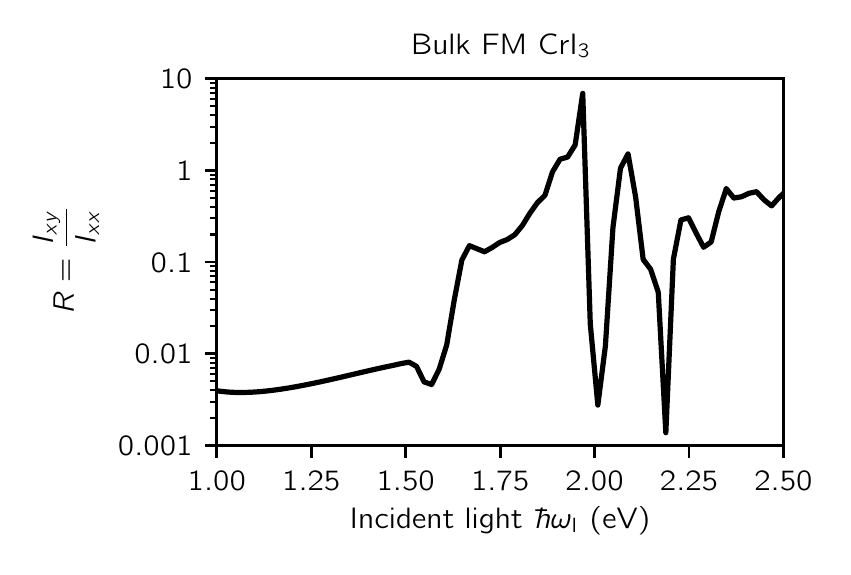}}
\caption{\label{fig:intensity1} The calculated ratio $R$ of mode $A_{\rm g}$ in bulk ferromagnetic CrI$_3$.  ($R$ is the ratio of blue and red curve from Fig.~\ref{fig:intensity}.)}
\end{figure}

\begin{figure}[h]
\centerline{\includegraphics{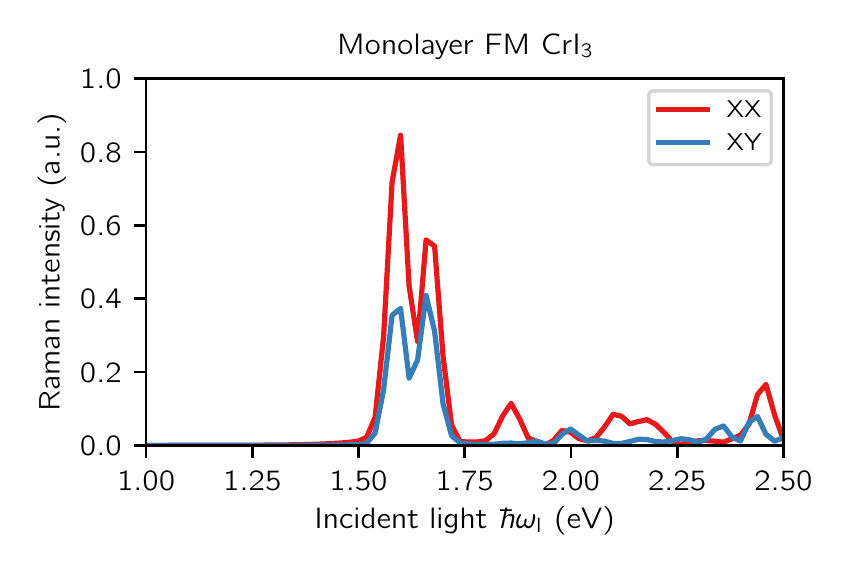}}
\caption{\label{fig:mono}The calculated Raman intensity of $A_{\rm g}$ mode in monolayer ferromagnetic CrI$_3$. The photon energy on the horizontal axis is shifted to reproduce the GW-BSE calculated optical bandgap onset.  The vertical scale here can't be compared with the bulk calculation from Fig.~\ref{fig:intensity} due to the reduced dimensionality of the monolayer as compared to bulk.}
\end{figure}

\begin{figure}[h]
\centerline{\includegraphics{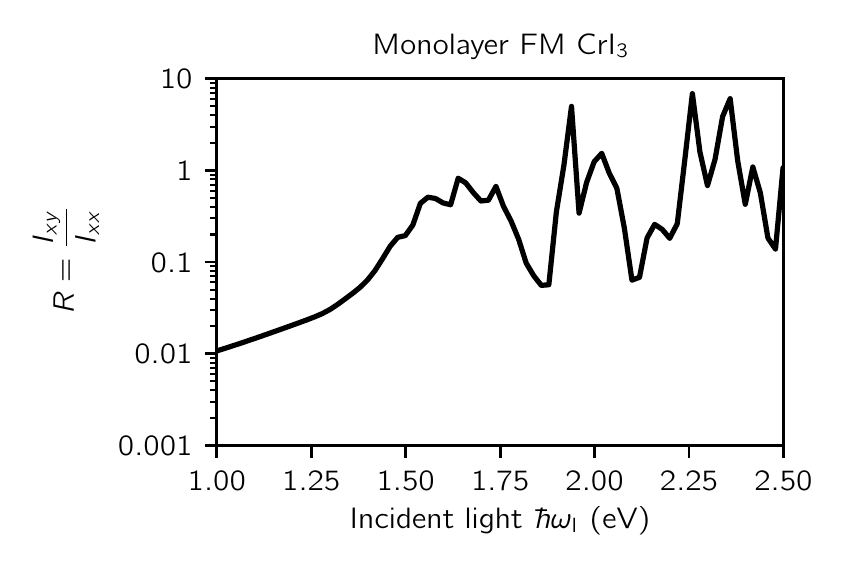}}
\caption{\label{fig:mono_ratio}
The calculated ratio $R$ of mode $A_{\rm g}$ in monolayer ferromagnetic CrI$_3$.  ($R$ is the ratio of blue and red
curve from Fig.~\ref{fig:mono}.)}
\end{figure}

Our main results for bulk CrI$_3$ are shown in Figs.~\ref{fig:intensity} and \ref{fig:intensity1}. We find that the intensity ratio $R$ varies greatly with the incoming laser energy.  When the energy of the incoming photon is below the band gap (1.5~eV) the ratio $R$ is close to 0.01.  At energies close to the band gap the ratio $R$ increases.  For example, when the incident photon energy is 1.8~eV, $R$ is about 0.5.  At somewhat larger incoming photon energy of 1.92~eV ratio $R$ is about 7. At even higher incoming photon energies, above 2.1~eV, the ratio $R$ fluctuates a lot and is on average around 0.5.

\subsection{Monolayer FM CrI$_3$}

Next, we discuss the calculated Raman intensities $I_{xx}$ and $I_{xy}$ of the 128~cm$^{-1}$ mode in monolayer CrI$_3$. The DFT computed monolayer bandgap is 0.82~eV, and the horizontal axis is shifted by $0.50$~eV so that it matches the GW-BSE calculated monolayer optical band gap of 1.32~eV.\cite{wu2019physical} From Fig.~\ref{fig:mono} we find, below the optical bandgap 1.32~eV, the intensities $I_{xx}$ and $I_{xy}$ are again small and their relative ratio is somewhat larger than in the bulk, as shown in Fig.~\ref{fig:mono_ratio}.

The main difference between monolayer and bulk occurs near the bandgap. In the monolayer both $I_{xx}$ and $I_{xy}$ start to increase at about 0.2~eV above the bandgap, while in the bulk $I_{xx}$ starts to increase already at the bandgap edge.  The different behavior of $I_{xx}$ in the bulk and monolayer can be explained by the band structure. Based on our band structure calculation, we find in the monolayer, both the valence band maximum (VBM) and conduction band minimum (CBM) are located at the $\Gamma$ point, while in the bulk, the VBM is at the $\Gamma$ point and CBM is at the T point.\cite{Kumar_Gudelli_2019} More importantly, the minimum direct bandgap in bulk is at the T point, rather than at the $\Gamma$ point, as in the monolayer. Clearly, matrix elements for transitions at these two points are different, which leads to different Raman signatures.  The transitions near the $\Gamma$ point activate mostly $I_{xx}$ while the ones near the T point activate both $I_{xx}$ and $I_{xy}$.

\subsection{Circularly polarized Raman}

Now we switch to the Raman intensity of mode $A_{\rm g}$ with circularly polarized incident light. The Raman tensor from Eq.~\ref{eq:tensor_mag} gives rise to the following Raman intensities for circularly polarized light,
\begin{align}
I^{\sigma^+ \sigma^+} &\sim  \left\vert a + i b \right\vert^2,\\ 
I^{\sigma^- \sigma^-} &\sim  \left\vert a - i b \right\vert^2, \\ 
I^{\sigma^+ \sigma^-} &= I^{\sigma^- \sigma^+} = 0.  \label{eq:circular}
\end{align}
Here $\sigma^+$ ($\sigma^-$) denote the left-handed (right-handed) circularly polarized incoming and outgoing light. Presence of an anti-symmetric $b$ term in Eq.~\ref{eq:tensor_mag} therefore implies that intensities $I^{\sigma^+ \sigma^+}$ and $I^{\sigma^- \sigma^-}$ are different from each other as long as $b\neq0$.  Furthermore, regardless of value of $b$ intensities $I^{\sigma^+ \sigma^-}$ and $I^{\sigma^- \sigma^+}$ remain zero.  (In contrast, a symmetric off-diagonal Raman tensor, as would be induced by a hypothetical breaking of a 3-fold axis (as in Eq.~\ref{eq:tensor_no3}), one would have $I^{\sigma^+ \sigma^+} = I^{\sigma^- \sigma^-} \sim |a|^2$, independent of $b$.  Furthermore, a symmetric off-diagonal component as in Eq.~\ref{eq:tensor_no3} would also imply that $I^{\sigma^+ \sigma^-} = I^{\sigma^- \sigma^+}\sim |b|^2$.)

\begin{figure}[h]
\centerline{\includegraphics{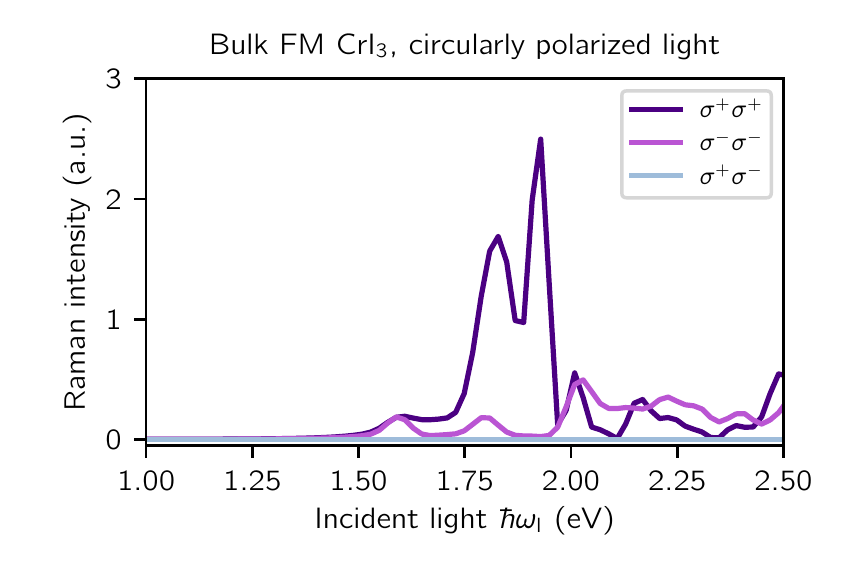}}
\caption{\label{fig:circular} The calculated circularly polarized Raman intensity of $A_{\rm g}$ mode in bulk ferromagnetic CrI$_3$. The photon energy on the horizontal axis is shifted to reproduce the GW-BSE calculated optical bandgap onset.}
\end{figure}

The calculated circularly polarized Raman intensity of bulk is shown in Fig.~\ref{fig:circular}.  We find when the magnetic moment is along the $+z$ direction, the Raman intensity is dominated by $\sigma^+ \sigma^+$ signal. For example, $I^{\sigma^+ \sigma^+}$ is more than 8 times larger than $I^{\sigma^- \sigma^-}$ when the incident light energy is between 1.65 and 1.95~eV. While when the magnetic moment flips from $+z$ direction to $-z$ direction, we confirmed that the Raman intensity is then dominated by $\sigma^- \sigma^-$ signal, as one would expect. A similar experimental result is obtained in Ref.~\onlinecite{huang2020tuning}: when the magnetization is along the $+z$ direction, $I^{\sigma^+ \sigma^+}$ is 3.8 times larger than $I^{\sigma^- \sigma^-}$ at the 128~cm$^{-1}$ A$_{\rm g}$ mode. By flipping the magnetization from $+z$ direction to $-z$ direction, the dominant Raman scattering changes from $\sigma^+ \sigma^+$ to $\sigma^- \sigma^-$ signal.

\begin{figure}[h]
\centerline{\includegraphics{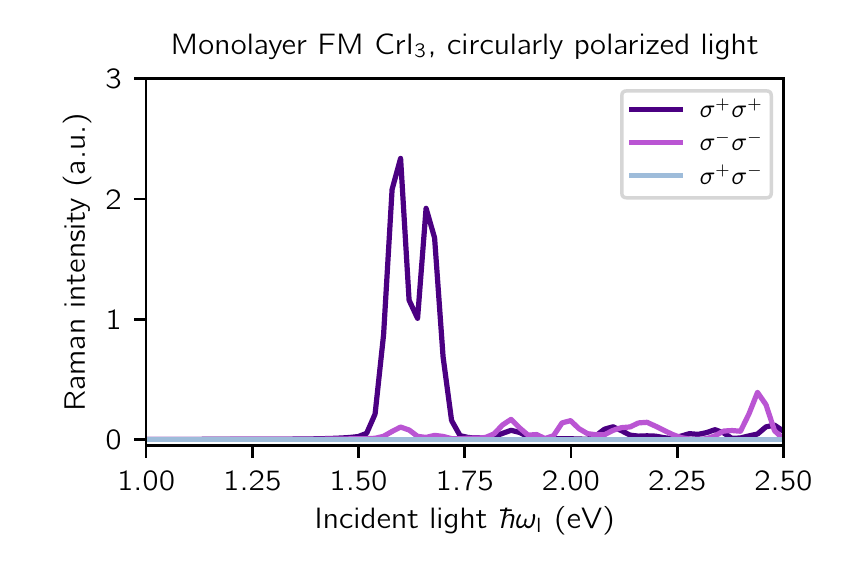}}
\caption{\label{fig:circular_mono}
The calculated circularly polarized Raman intensity of $A_{\rm g}$ mode in monolayer ferromagnetic CrI$_3$. The photon energy on the horizontal axis is shifted to reproduce the GW-BSE calculated optical bandgap onset. The vertical scale here can't be compared with the bulk calculation from Fig.~\ref{fig:circular} due to the reduced dimensionality of the monolayer as compared to bulk.}
\end{figure}

Furthermore, we computed for completeness the circularly polarized Raman intensity of monolayer, as shown in Fig.~\ref{fig:circular_mono}. The main difference with respect to the bulk occurs near the band gap edge, which we again contribute to the different nature of optical transitions at $\Gamma$ and T points.

\subsection{Electron lifetime}
\label{sec:lifetime}

\begin{figure}[h]
\centerline{\includegraphics{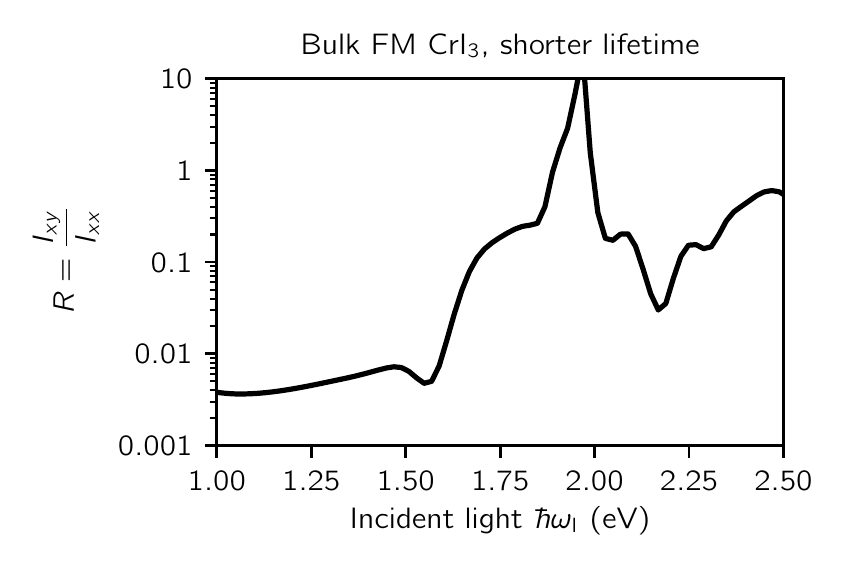}}
\caption{\label{fig:smearing} The calculated ratio $R$ in bulk ferromagnetic CrI$_3$ using a shorter electron lifetime, $\delta$=0.16~eV. In Fig.~\ref{fig:intensity1} we used $\delta$=0.10~eV.}
\end{figure}

Now we investigate the effect of electron lifetime on the Raman intensity. The Raman intensity ratio $R$ with shorter electron lifetimes ($\delta$=0.16~eV) is shown in Fig.~\ref{fig:smearing}. Comparing with Fig.~\ref{fig:intensity1} we find that a shorter electron lifetime doesn't change the overall ratio $R$, but that it only dampens oscillations of $R$ with the energy of light $\omega_{\rm I}$.  Therefore, the effect of increased $\delta$ is the overall reduction in both $I_{xx}$ and $I_{xy}$, so that $R$ is not changed much.

\subsection{Magnetic moment}

Now we analyze the Raman signature $I_{xy}$ of all four A$_{\rm g}$ modes from the point of view of the magnetic moment. In the limit of a weak spin-orbit coupling, 
the susceptibility $\chi_{xy}$ is linearly proportional to the atomic magnetic moment $m$ on Cr.  Therefore, one might expect that
the derivative of the susceptibility $\frac{\partial \chi(\omega_{\rm I})}{\partial u}$ will also be proportional to the derivative of the magnetic moment, $| \frac{\partial m}{\partial u} |$. 
Therefore, atomic displacements that change the magnetic moment more should also give rise to a larger off-diagonal Raman intensity.
To check this assumption, we computed $\frac{\partial m}{\partial u}$ for all four A$_{\rm g}$ modes.  First, we calculate the magnetic moment of a single Cr atom in the ground state, without any atomic displacements. Second, we displace atoms by a small finite amount ($u$) and recalculate the magnetic moment of the Cr atom.  This difference then gives us an estimate of $\frac{\partial m}{\partial u}$.  We repeated this calculation with a range of atomic displacement magnitude ($u$) to confirm the linear change of $m$ with $u$. 

For the displacement of I atoms along the $\pm z$ axis we find that $|\frac{\partial m}{\partial u}|$ equals $0.98$~$\mu_{\rm B}/\rm{\AA}$.  As expected, $|\frac{\partial m}{\partial u}|$ is smaller for the remaining three A$_{\rm g}$ modes.  Specifically, for the displacement of Cr atoms along the $\pm z$ direction we find that $|\frac{\partial m}{\partial u}|$ is $0.27$~$\mu_{\rm B}/\rm{\AA}$, while for the two modes corresponding to I atom displacement in the $x$--$y$ plane $|\frac{\partial m}{\partial u}|$ is only $0.06$ and $0.04$~$\mu_{\rm B}/\rm{\AA}$.  Comparing calculated $|\frac{\partial m}{\partial u}|$ with averages of $\frac{\partial \chi (\omega_{\rm I})}{\partial u}$ defined in Eq.~\ref{eq:averageschi} we indeed find that the mode that changes the magnetic moment the most has the largest off-diagonal Raman signature.

Interestingly, the dominant displacement modulating magnetic moment is coming from the displacement of nominally non-magnetic iodine atom, not from magnetic chromium atom.  We discuss this in more detail in Sec.~\ref{sec:spin-orbit} where we demonstrate that the off-diagonal raman signature here originates dominantly from the spin-orbit interaction of iodine atom.

\subsection{Phonon modulated MOKE}

\begin{figure}[h]
\centerline{\includegraphics{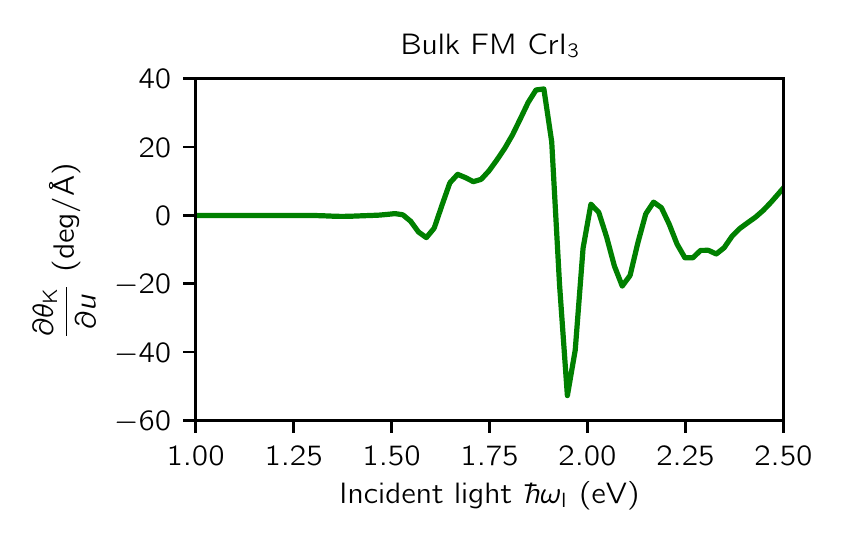}}
\caption{\label{fig:kerr} The calculated phonon modulated MOKE signal in bulk ferromagnetic CrI$_3$.}
\end{figure}

Motivated by the large magnitude of $\frac{\partial m}{\partial u}$ we decided to compute the phonon modulated magneto-optical Kerr effect (MOKE), as it is often used as a signal of the magnetic moment in magnetic materials.\cite{Guo_2018, Kumar_Gudelli_2019, wu2019physical, huang2017layer} We calculated the Kerr angle $\theta_{\rm K}$ using,\cite{alma, PhysRev.186.891}
\begin{align}
\theta_{\rm K} = - {\rm Re}\left[\frac{\epsilon_{xy}}{\sqrt{\epsilon_{xx}}\left(\epsilon_{xx}-1 \right)}\right].
\end{align}
We again displaced atoms and recomputed change in the Kerr angle $\theta_{\rm K}$ to obtain $\frac{\partial \theta_{\rm K} (\omega_{\rm I}) }{\partial u}$. The results for bulk CrI$_3$ are shown in Fig.~\ref{fig:kerr}.

We find numerically that $\frac{\partial \theta_{\rm K}}{\partial u}$ can be well approximated with,
\begin{align}
\approx - {\rm Re}\left[
\frac{1}{\sqrt{\epsilon_{xx}}\left(\epsilon_{xx}-1 \right)}
\frac{\partial \epsilon_{xy}}{\partial u}
\right].
\end{align}
Therefore, we conclude that in our case $\frac{\partial \theta_{\rm K}}{\partial u}$ is approximately proportional to the coefficient $b$.  Therefore, phonon modulated MOKE signal can be directly compared to the cross-polarized Raman signature $I_{xy}$, which is proportional to $|b|^2$.  In particular, we find that the large positive peak in phonon modulated MOKE ($\sim 40$~deg/\AA\ at 1.86~eV) and large negative peak (about $-55$~deg/\AA) at somewhat higher energy  (1.93~eV) are in direct correspondence with two positive-definite peaks in the $I_{xy}$ signature shown in blue in Fig.~\ref{fig:intensity}.

\section{Analysis}
\label{sec:analysis}

The presence of $I_{xy}$ signal in CrI$_3$ depends on three ingredients, as discussed earlier: broken time-reversal in ferromagnetic state, spin-orbit, and resonance.  Now we analyze these conditions in more detail, one by one.

\subsection{Anti-ferromagnetic order}

First to further confirm the importance of the ferromagnetic order for non-zero $b$, we computed Raman intensity in the antiferromagnetic (AFM) state of CrI$_3$. If we assume that FM and AFM states have the same crystal structure, but that only the atomic magnetic moments point in different directions, we infer that the magnetic space group of AFM state is R$\overline{3}'$.  This space group consists of a 3-fold rotation, as well as inversion followed by time-reversal.  The inversion operation, or time-reversal operation, on their
own are not symmetries in the AFM state.
\begin{figure}[H]
\centerline{\includegraphics{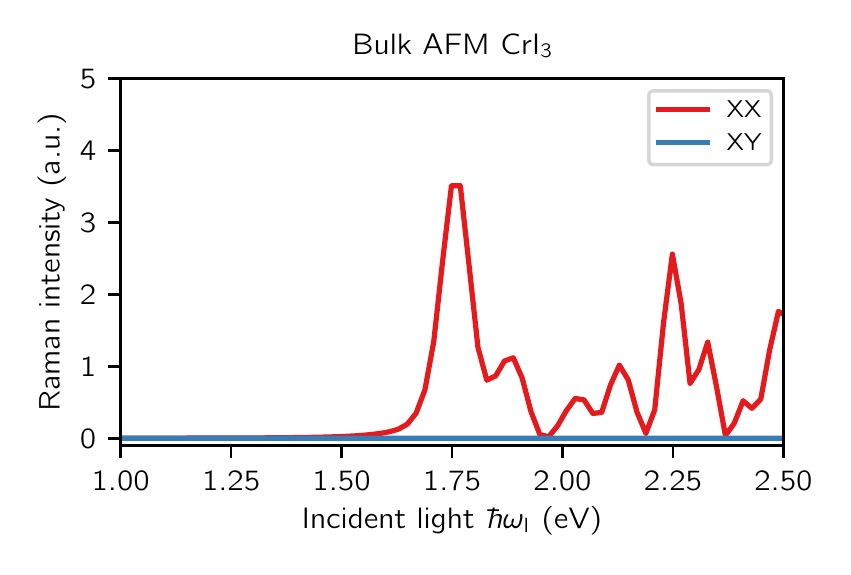}}
\caption{\label{fig:AFM} The calculated Raman intensity of $A_{\rm g}$ mode in bulk CrI$_3$ in the anti-ferromagnetic state, showing no $I_{xy}$ intensity.}
\end{figure}
Therefore the symmetry analysis in this case gives that the Raman signature should not have the cross-polarized signal $b$ and that therefore $R=0$.  This is exactly what we find in our calculation, as shown in Fig.~\ref{fig:AFM}. While the $I_{xx}$ intensities are of similar structure in the AFM and FM state, we find that the $I_{xy}$ intensity is exactly zero in the AFM state, as expected from the symmetry arguments.

\begin{figure}[h]
\centerline{\includegraphics{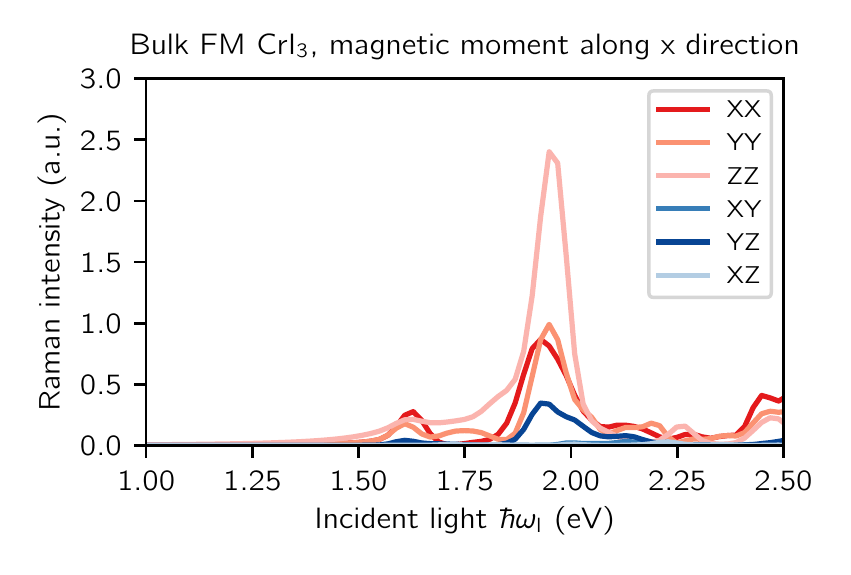}}
\caption{\label{fig:xyplane} The calculated Raman intensity of
$A_{\rm g}$ mode in bulk ferromagnetic CrI$_3$ with magnetic moment pointing along the in-plane $x$-axis direction, showing large cross-polarized $I_{yz}$ intensity.}
\end{figure}

Furthermore, we investigate the Raman intensity in the FM state when the magnetic moment is along the $x$ direction. Walker\cite{mccreary2019distinct} found when the applied magnetic field was parallel to the monolayer and the magnetic field was large enough ($B=6$~T), the CrI$_3$ layers had a magnetic moment in the $x$--$y$ plane. Therefore, we decided to repeat our calculation with the magnetic moment along the $x$ direction, and the calculated Raman intensity of mode $A_{\rm g}$ is shown in Fig.~\ref{fig:xyplane}. We find the $I_{xy}$ intensity is almost zero, and we observe a strong $I_{yz}$ intensity which is comparable to $I_{xx}$. For example, when the incoming photon energy is 1.91~eV, $I_{yz}$ is about half of $I_{xx}$ and ten times larger than $I_{xy}$. 

\subsection{Spin-orbit interaction}
\label{sec:spin-orbit}

\begin{figure*}[htbp]
\centerline{\includegraphics{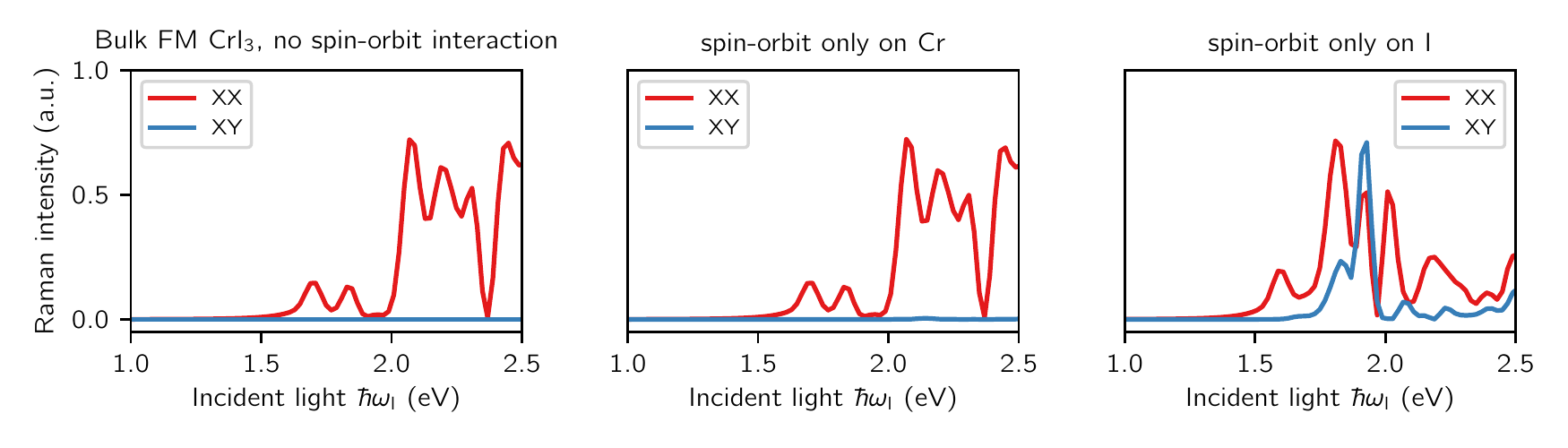}}
\caption{\label{fig:soc} The calculated Raman intensity of mode $A_{\rm g}$ in the bulk ferromagnetic CrI$_3$. The left panel shows result without spin-orbit interaction. The middle panel shows the result with spin-orbit interaction on only Cr atoms. The right panel shows the result with spin-orbit interaction on only I atoms (nearly the same as that in a calculation where spin-orbit is included on all atoms, see Fig.~\ref{fig:intensity}.)}
\end{figure*}

Second, we consider the importance of the spin-orbit interaction.  We confirmed that if spin-orbit is excluded from the calculation that we obtain $b=0$, as expected.  However, we can take this analysis one step further.  The pseudopotentials for Cr and I in our calculation were constructed by solving the relativistic Dirac equation separately on Cr and I.  Therefore, this gives us a possibility to perform a calculation where spin-orbit is turned on selectively on either Cr or I atom.  The results of such calculations are shown in Fig.~\ref{fig:soc}. As can be seen from the middle panel of the figure, if spin-orbit interaction is included only on Cr atoms we get nearly negligible $I_{xy}$ intensity.  This is a somewhat surprising result, as one might naively assume that locality of the spin-orbit interaction potential implies that the only relevant spin-orbit interaction is the one on the magnetic atom.  Clearly, there must therefore be some significant spillover of the dominantly chromium atomic-like orbitals onto iodine atoms. 

Given this insight we then repeated calculation with spin-orbit interaction present only on the iodine atom (right panel of Fig.~\ref{fig:soc}).  We find that the calculated Raman signature is nearly indistinguishable from that in the full Raman calculation.  Therefore, we conclude that the spin-orbit on iodine atom is almost exclusively responsible for calculated $I_{xy}$ intensity.  This observation is again consistent with our earlier finding that the largest change of the magnetic moment ($\frac{\partial m}{\partial u}$) comes from the displacement of I atom, not Cr atom.   Finally, our findings are also consistent with the results in Refs.~\onlinecite{PhysRevB.99.104432, Tartagliaeabb9379, Lado_2017, D0TC01322F, Mukherjee_2019}.

\subsection{Resonance}

Third, we discuss the importance of resonance.  As can be seen from Fig.~\ref{fig:intensity1} the $I_{xy}$ intensity is already two orders of magnitude smaller than $I_{xx}$ at low energy.  We find that $I_{xy}$ vanishes in the $\omega_{\rm I}=0$ limit, as is expected in an insulator based on our earlier analysis.  Therefore, without taking into consideration $\omega_{\rm I}$ dependence of $\chi$ the $I_{xy}$ intensity would be zero.

\subsection{Relevant atomic-like orbitals}

The Wannier interpolation\cite{MOSTOFI20142309} method that we used to compute optical response allows us to analyze the results in more detail.  For this purpose, we analyzed the exact tight-binding parameters we obtained from Wannier interpolation.  If we look at the on-site and hopping energies in the case of undistorted CrI$_3$, we find that the dominant energy scale near the Fermi level is the splitting of the minority and majority spin states on Cr atoms. Therefore, we can for simplicity consider only the majority spins, as the gap to the minority spins is about 2~eV.  Focusing on the five majority spin $d$-like states we find that they are split into three t$_{2\rm g}$-like states and two e$_{\rm g}$-like states, as expected in the octahedral environment. The splitting between  t$_{2\rm g}$ and e$_{\rm g}$ states is about 0.5~eV.  The electron bands formed by the t$_{2\rm g}$-like states are fully occupied, while e$_{\rm g}$-like bands are empty, as would be expected for spin-polarized nominally Cr$^{3+}$ state in CrI$_3$.

Next we displace the I atoms along the $\pm z$-axis, corresponding to the A$_{\rm g}$ phonon mode.  If we repeat our Wannier analysis on the distorted structure we find that this displacement of I atoms dominantly results in modulation of the crystal field splitting between t$_{2\rm g}$ and e$_{\rm g}$ states. This is to be expected from a simplified crystal field theory, as this particular mode modulates all Cr--I bond lengths in the octahedron. The calculated magnitude of the splitting is about $2$~eV/\AA.  We also checked that the onsite energies are not changed as much.  The largest change in the hopping term is $-0.9$~eV/\AA. 

\begin{figure}[h]
\centerline{\includegraphics{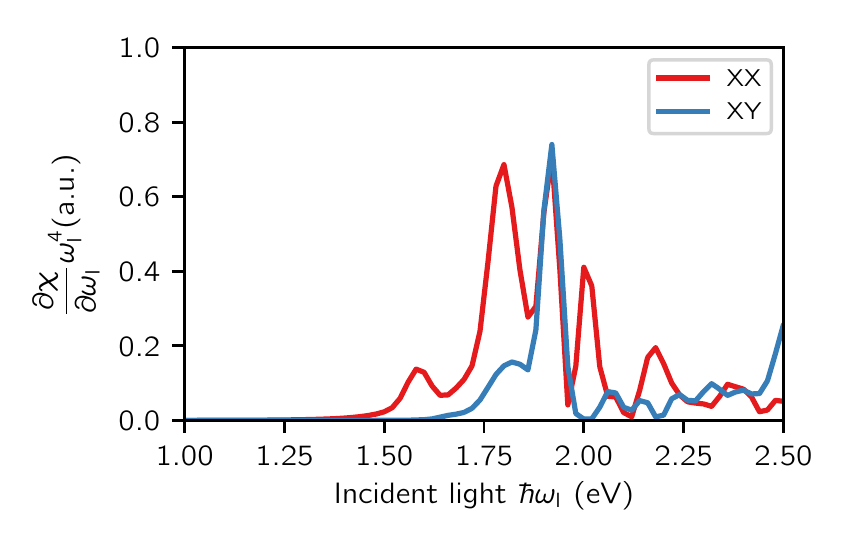}}
\caption{\label{fig:displacement} Approximate Raman intensity one would get by assuming that dominant atomic motions modulate the t$_{2 {\rm g}}$--$e_{\rm g}$ crystal field splitting in CrI$_3$.  See text for details. The photon energy on the horizontal axis is shifted to reproduce the GW-BSE calculated optical bandgap onset.}
\end{figure}

Therefore, the dominant role of the displacement of I atoms along the $\pm z$-axis is to rigidly shift the unoccupied e$_{\rm g}$ bands relative to the occupied t$_{2 \rm g}$ bands. Therefore, if we consider derivative of Eq.~\ref{Kubo} with respect to atom displacement $u$, we find that in this simplified picture of rigid displacement of bands, the following approximation holds,
\begin{align}
I \sim \left \vert \frac{\partial \chi}{\partial u} \right \vert^2 
\sim
\left \vert
\frac{\partial \chi}{\partial \omega_{\rm I}}
\right \vert^2. \label{displace}
\end{align}
We confirmed this by an explicit numerical calculation. Figure~\ref{fig:displacement} shows calculated  $\left \vert \frac{\partial \chi}{\partial \omega_{\rm I}} \right \vert^2  \omega_{\rm I}^4$ as a function of photon energy.  As can be seen from the figure, the Raman intensity is in good qualitative agreement with full calculation, where intensity is computed from $\frac{\partial \chi}{\partial u}$ (see Fig.~\ref{fig:intensity}). As one would expect, the agreement is better in the region just above the band gap edge. Therefore, we conclude that the main origin of the Raman signature in our calculations is the rigid shifts in the band structure due to the displacement of iodine atoms along the $\pm z$-axis.  

\section{Conclusion}
\label{sec:conclusion}

In conclusion, our calculations show strong Raman scattering in CrI$_3$ driven by ferromagnetism, spin-orbit interaction, and resonance effects.  The strong $I_{xy}$ polarized Raman scattering comes from the spin-orbit effect of I atoms, rather than the metal Cr atoms. We speculate that our findings open up the possibility that strong spin orbit effects in materials need not come from the magnetic ion, but could instead be driven by non-magnetic ion, as in this case. Therefore, it might be possible to find other materials with strong spin-orbit interaction, even when magnetism originates from atoms with small spin-orbit interaction, such as Cr, Mn, Fe, Co, or Ni.

\acknowledgements{This work was supported by grant NSF DMR-1848074. Computations were performed using the HPCC computer cluster at UCR.}

\bibliography{pap}

\end{document}